\documentstyle[12pt]{article}
\newcommand{\bn}{\begin}
\newcommand{\be}[1]{\begin{equation}\label{#1}}
\newcommand{\ee}{\end{equation}}
\newcommand{\bd}[1]{\begin{displaymath}\label{#1}}
\newcommand{\ed}{\end{displaymath}}
\newcommand{\ci}[1]{\cite{#1}}
\newcommand{\lb}[1]{\label{#1}}

\newcommand{\bea}{\begin{eqnarray}}
\newcommand{\eea}{\end{eqnarray}}
\newcommand{\eref}[1]{{(}\ref{#1}{)}}   %  equation ref
\title{Notes on the Generalised Second Law of Thermodynamics}
\author{S.--T. Sung\thanks{E-mail address: s.t.sung@durham.ac.uk}\\
        Department of Mathematical Sciences\\
        University of Durham, Durham DH1 3LE, U.K.}
\begin{document}
\maketitle
%
%
%
%
%       -----  abstract   ---------------
%
%
%
\begin{abstract}

Several comments are given to previous proofs of the generalised second law of thermodynamics: black hole entropy plus ordinary matter entropy never decreases for a thermally closed system. Arguments in favour of its truism are given in the spirit of conventional thermodynamics. 

\end{abstract}
%\pagebreak
%
%
%
%        -------   TeXt ---------------------
%
%     --------- One Section Only   --------------
%
%
%\bs{Introduction}{intro}
%\baselineskip=3.8ex

Thermodynamics (see \ci{cal85,hua87} for introduction) is one of the physical disciplines in which physical laws are governed by simplicity and generality (S\&G). Due to the largeness of physical degrees of freedom, most macroscopic systems are untraceable microscopically. Therefore, a systematic way, based on macroscopic S\&G regardless of the microscopic details, is needed in order to extract the information we are interested, amongst which one of the most important is perhaps the equilibrium states. Thermodynamics suffices such task by employing extremising (maximising or minimising) principles \ci{cal85}. Even though a microscopic model is introduced later to give thermodynamic quantities a statistical-mechanical interpretation, the major roles of thermodynamic quantities, and hence the extremising principles, are unquestionable.
 
Like all other microscopic physical laws, the status of second law of thermodynamics (SLT) is more a postulate than a theorem \ci{cal85,hua87}: It is taken as one of the starting points for a long journey of searching statistical descriptions of a physical system. Therefore, it has to be checked up again and again throughout the journey. In other words, it can only be verified in a {\it self-consistent} way, or in a {\it circular} way by Callen's word \ci{cal85}. And because SLT is an experience law applicable only at macroscopic scale, it cannot survive under the closest scrutiny from the viewpoint of microscopic unitary evolution. Even so, up to now, we have all of the reasons to believe that its S\&G is unquestionable if we do not go beyond the border.
% See Callen's page 15

On the other side of physics, we are used to thinking of space-time in geometric language after Einstein formulated general relativity, which is always regarded as a dynamic theory. Presumably, it would have surprised him very much, as we are, that, along with the development of black hole physics \ci{haw73,wald84}, gravitational degrees of freedom can also be cast in the language of thermodynamics \ci{jac95}, as can be seen most transparently from the identification of the area of an event horizon with entropy \ci{bek73b,haw75} and the formulation of four laws of black hole mechanics \ci{bar73}.

However, as being pointed out by Callen \ci{cal85}, thermodynamics by itself is not a theory; it is a way of thinking: thinking about the laws of nature which are universal and revealed in macroscopic scale whatever the microscopic compositions and dynamics the system has. From this point of view, it should not surprise us anymore that gravity, which is usually neglected in thermodynamics because of its weakness, can/should also be incorporated into thermodynamics.

One of the most important developments along this line is Bekenstein's conjecture \ci{bek73b} that the second law of thermodynamics and that of black hole mechanics should indeed be combined together as a generalised second law (GSL) for a closed self-gravitating system. It states that for a gravitationally closed system which is consisted of a black hole and ordinary matter, the total entropy should never decrease. 

Bekenstein's conjecture is proposed before Hawking's celebrated discovery that a black hole is not only a big sucker, but also a blood-giver which can radiate. Therefore, at pre-Hawking radiation era, the GSL is only an approximated thermodynamic law which can be violated provided that the black hole is surrounded by thermal radiation of temperature lower than that of the black hole. Hawking's discovery gives the GSL a chance of surviving. Once more, we can expect that S\&G still governs thermodynamic laws.

After Bekenstein offered his conjecture of the GSL, strong evidences for its truism have been given in references \ci{bek73b}, \ci{bek74a}--\ci{zas96}. It is thus natural to regard the GSL as a special case of the SLT which involves a black hole.
%,unr82,zur82,sch85,zur85,fro93,

Nonetheless, with above attitude towards the SLT in mind, we feel that the {\it proofs} of the GSL available to us are unsatisfactory on two aspects: The first, the status of the GSL in thermodynamics is not revealed explicitly. As being stressed above, the GSL of thermodynamics is not a consequence of any other physical laws within thermodynamics (or statistical mechanics); it is the starting point of the following story: By maximising entropy we can determine the equilibrium states of the system. It is virtually hopeless to do this following the microscopic dynamic evolution. Conversely, as far as we know, only an equilibrium state has well-defined thermodynamic functions of state; entropy is one of them. The second, the flavour of thermodynamics---simplicity and generality---is veiled by the detailed microscopic dynamics.    
   
In this note, we would like to offer examples to show how the GSL works and evidences for its truism from the point of view described above. Because we have accepted its truism as the {\it first law}, the arguments is {\it not} a proof, but self-consistent statements that serves as its foundation. 

Before we present our approach, we would like to give a few comments to previous attempts of proof. Although Frolov and Page have given several comments to those prior to theirs \ci{fro93}, we would like to add some to contrast those approaches with our attitude.

Since Bekenstein offered his conjecture before Hawking radiation was discovered, his proof \ci{bek73b,bek74a} suffered from the unavoidable incompleteness in which the entropy of radiation was missed. He proposed a lower bound of spatial expansion (with respect to fixed $S$ and $U$) for ordinary thermodynamic system as a remedy. We think the generalised second law can be understood without introducing such a bound (as shown later). On the other hand, nor is it sufficient to guarantee the validity of the generalised second law: Consider the case in which the initial and final masses of the black hole are the same, then the entropy difference comes purely from those matter outside the black hole. Due to the universality of Bekenstein's bound \ci{bek94}, it cannot inform us how to calculate this difference.       

In Unruh and Wald's version \ci{unr82} (see also Zaslavskii's \ci{zas96}), the importance of the entropy contribution from radiation was stressed, and it was used to remedy the incompleteness of Bekenstein's proof by considering buoyancy force originated from radiation. This buoyancy force will be felt by matter in a stationary Schwarzschild frame (SSF), but not in a locally inertial frame (LIF). As a consequence, after the rope is cut, which ties matter to someone standing outside the system concerned, such buoyancy force can be forgotten. On the other hand, because the GSL concerns the never-decreasing property of entropy of a thermally closed system, what we are interested is the entropy change during the period of approaching equilibrium {\it after} the rope is cut when the system can be regarded as thermally closed. Nevertheless, the non-negligibility of the entropy of Hawking radiation will never be over-stressed. We will see later that the GSL is rescued not only by the existence of Hawking radiation, but also by its massless and thermal properties.

Zurek's proof \ci{zur82}(later generalised by Schumacher \ci{sch85}) gave a strong support to Bekenstein's statistical interpretation of black hole entropy as the lack of information about the internal configurations of a black hole. Nonetheless, his proof only achieved half of the goal: if the surrounding thermal radiation has a higher temperature than that of the black hole, the GSL will be violated. However, as we will show later, if the equilibrium states can be handled properly, a final equilibrium state with higher entropy can always be found. 

As a side product, the GSL was derived within the membrane paradigm by Zurek and Thorne \ci{zur85}. In fact, they used the thermal atmosphere to transform the statement of the GSL to a special case of the SLT. Nonetheless, the importance and usefulness of the GSL will be gravely weakened if it has to be understood with the help of FIDOs near an event horizon from a local point of view because then, it becomes unclear how the GSL can be used globally to the whole system (this is how we use the SLT and this is also the reason why we need thermodynamics). We prefer doing without FIDOs' help.

In Frolov and Page's proof \ci{fro93}, quantum modes in the interior of a black hole were assumed to be the CPT reversal of those outside a black hole. We think this assumption is so rigid that once it is invalidated, the whole arguments will break down. A different picture of the interior of black holes could indeed be drawn \ci{sung97a}. And since the black hole entropy can be determined at spatial infinity, it seems unnecessary to worry about what are happening inside a black hole. On the other hand, in the last step of their proof, they compared two Massieu functions defined on different spaces---exterior and interior of the black hole. However, entropy is an extensive quantity (this is still {\it roughly} true for a composite system which is not separated by an impermeable wall), its value depends on the volume where the system is confined. In the terminology of quantum field theory, it depends on the normalisation. It is thus unclear, since the practical computation scheme was not given in such a general approach, whether their difference can be explicitly calculated to be semi-positive definite, as claimed in their proof.

To help clarify the scenario of our approach, let us review the basic statement of the SLT at first. From the point of view of entropy, it says that the entropy ($S$) for a thermally closed system ($\delta Q=0$) will never decrease. For a thermally closed system in equilibrium, {\it by definition} (since this is how we determine the equilibrium state), $S$ is maximum, and the temperature $T=1/\beta$, defined as $\beta=({\partial S}/{\partial U})_{V}$ (we employ units with $\hbar=c=G=k_B=1$), is a constant throughout the system.

How can we benefit from the SLT? Consider a composite system which is consisted of two boxes of matter attached to each other along a wall which is restrictive with respect to energy (ERWall)\ci{cal85}; and each one is in equilibrium by its own. Then, we change the wall to a heat-permeable (or particle-permeable) one with negligible effects on the system so that they can approach final equilibrium state. The final state is determined by maximising the entropy with internal energy and volume keeping fixed. Even though this is not the only way to get the information about the final state, this is perhaps the simplest one. In this case, the SLT is verified directly from daily experiences.  

We will apply a similar picture to the GSL. As mentioned previously, the GSL will be violated without Hawking radiation. Consequently, we need to include Hawking radiation to form an equilibrium state involving a black hole. We confine a black hole ($H$, we consider Schwarzschild black holes only) and thermal radiation ($R$) of total internal energy (ADM mass) $U_i=M_{Hi}+U_{Ri}$ at temperature $T_i=1/\beta_i$ in a ball ($B$) of volume $V$ (black hole has no volume by prescription). This picture was pioneered by Hawking \ci{haw76a}. (This picture contains the phenomenon of phase transition in which $M_H$ acts as an order parameter \ci{hua87}. Black hole formations as phase transitions have been observed in numerical simulations \ci{cho93,abr93,eva94}. Though those simulations are done in a purely classical setting, we believe those results can still be regarded as a partial support for the applicability of our picture.)

Some remarks are needed since $V$ and $U_{Ri}$ will diverge in the SSF without proper prescriptions. Consider, at spatial infinity, a thin spherical shell of box of volume $\delta V=a\delta r$ (where $a$ is the area orthogonal to the radial co-ordinate and $\delta r$ is the radial expansion in the co-ordinates of the SSF) containing thermal radiation of temperature $T$. The entropy is then $S_{\delta}=\beta(U_{\delta}-F_{\delta})$, where $F_{\delta}$ is the corresponding Helmholtz free energy. If we put this shell at co-ordinate $r$, to an observer in a LIF, the energy and temperature then scale in the same manner by a factor $\chi=(1-2M/r)^{-1/2}$ \ci{tol34}. Therefore, the entropy is independent of frames (SSF or LIF) even though $\chi$ diverges at the event horizon. Furthermore, if we use a box with the same $\delta r$, then the energy density is also frame-independent since the volume also scales by a factor of $\chi$ (but entropy density is thus zero, and Stefan's constant is {\it not} a fundamental constant). We therefore build up the ball, shell by shell, at spatial infinity at first, then we pull it to the nearby of the black hole while at the same time keeping the radial co-ordinate expansion fixed. Since in our equations only entropy, hence the combinations of $\beta U$ and $\beta F$, will appear, we can dismiss factor $\chi$ totally. We thus can use these quantities as if we are in spatial infinity. This perhaps is one of the reasons that entropy is more important and interesting than other quantities.

The criteria for the existence of various configurations of the system described above have been analysed in \ci{haw76a,gib78a}. In our approach, these criteria are not used explicitly because it is not necessary to require that there is a black hole in either the initial or the final state.

To see how the GSL works, we attach to the outer surface of $B$ a spherical shell of box ($b$) of volume $v$ which can contain any kind of ordinary matter ($m$) at temperature $T_{\bar\imath}$ with internal energy $U_{\bar\imath}$ and Helmholtz free energy $F_{\bar\imath}$.  

From here, we can have two different approaches. The first, $B$ and $b$ are always separated by a heat-permeable wall while they are approaching equilibrium, so there is only heat exchange between them. The final matter form in $b$ is still $m$. The second, the wall between $B$ and $b$ will be removed so that $m$ will fall into the hole (if there is one), thus the final matter form in $b$ is thermal radiation ($r$). The second case shows a new feature of thermodynamics involving a black hole in which the black hole acts as a matter-to-radiation transformer (if $m$ is not thermal radiation in the first place). We consider these two cases separately.

From the first case we will learn how the GSL works and this will provide us with a basis for the consideration of the second case. The whole entropy change can be separated into three parts,
\be{e10}
dS=dS_H+dS_R+dS_{m}~,
\ee
where 
\bea
dS_H&=&4\pi M_{Hf}^2 -4\pi M_{Hi}^2~,\lb{e11.11}\\
dS_R&=&\beta_f (U_{Rf}-F_{Rf})-\beta_i (U_{Ri}-F_{Ri})~,\lb{e11.12}\\
dS_{m}&=&\beta_f(U_{mf}-F_{mf})-\beta_{\bar\imath}(U_{m{\bar\imath}}-F_{m{\bar \imath}})~.\nonumber
\eea
$U_{ab}$ and $F_{ab}$ are internal energy and Helmholtz free energy for matter form $a$ at temperature $T_b$, respectively.

To understand why the final state of the triple-phase system (if $M_{Hf}\not= 0$) has the highest entropy without doing calculation, let us consider the following slow-motioned thought experiment in which the whole system approaches equilibrium through an infinite-step procedure: We cover the black hole by an ERWall at first, then let $R$ and $m$ approach equilibrium with $U_R+U_m$ fixed (observed at spatial infinity). According to the SLT, the entropy change is semi-positive definite. Afterwards, we remove the ERWall around the black hole, but cover $B$ with an ERWall. Then, let $H$ and $R$ approach equilibrium with $M_H+U_R$ fixed (also observed at spatial infinity). According to the GSL, the entropy change is also semi-positive definite. We then carry on above procedure again and again until the whole system, $H$+$R$+$m$, arrives at equilibrium. Because entropy is a function of state, the entropy change is unique for a thermally closed system if so is the final state determined by maximising entropy. (We assume it is.) Therefore, the total entropy change is semi-positive definite. (The reader may suspect that with delicate arrangement, the above procedure could have no definite final state, just like the series $1,-1,1,-1,\ldots$ has no limit. However, it is not difficult to convince oneself that if $T_{\bar\imath}>T_i$ ($T_{\bar\imath}<T_i$), then $T$ of $m$ will decrease (increase) only. And if a phase transition happens, i.e., $M_{H}\not= 0\rightarrow M_{H}=0$ or $M_{H}= 0\rightarrow M_{H}\not=0$, then the other direction will not happen. Therefore, we can safely expect that the final equilibrium state will be the limit state of the above procedure.) Alternatively, one can write down those entropy terms explicitly and maximising the total entropy.

The second case is generic to the GSL because it involves transformations between different matter forms. The total entropy change can be separated into four parts,
\bd{e11.1}
dS=dS_H+dS_R+dS_r+dS_m~,
\ed
where $dS_H$ and $dS_R$ are respectively as those in \eref{e11.11} and \eref{e11.12}, and
\bea\lb{e11.2}
dS_r&=&\beta_f (U_{rf}-F_{rf})-\beta_{\bar{\bar\imath}}(U_{r{\bar{\bar\imath}}}-F_{r{\bar{\bar\imath}}})~,\nonumber\\
dS_m&=&\beta_{\bar{\bar\imath}}(U_{r{\bar{\bar\imath}}}-F_{r{\bar{\bar\imath}}})-\beta_{\bar\imath}(U_{m{\bar\imath}}-F_{m{\bar \imath}})~.\nonumber
\eea
$U_{r{\bar{\bar\imath}}}$ and $\beta_{\bar{\bar\imath}}=1/T_{\bar{\bar\imath}}$ are determined by replacing initial matter $m$ in $b$ with thermal radiation $r$ of identical internal energy, namely, $U_{r{\bar{\bar\imath}}}=U_{m{\bar\imath}}$. Now, if we consider a system with initial state $H_i+R_i+r_{\bar{\bar\imath}}$, then we can borrow the conclusion of the first case that the entropy change, $dS_H+dS_R+dS_r$, is semi-positive definite. However, if the final state of our original system $H_i+R_i+m_{\bar\imath}$ contains a black hole (i.e., $m$ will be swallowed by $H$), then this final state is just the same final state arrived from the initial state $H_i+R_i+r_{\bar{\bar\imath}}$. Therefore, if $dS_m$ is semi-positive definite, we then arrive at the desideratum. Though it is quite unlikely to give a proof to this statement, it seems intuitively true. On the other hand, if the final state of $H_i+R_i+m_{\bar\imath}$ does not contain a black hole, then we come back to the first case.

From the other point of view, by accepting the truism of GSL in the first place (as we did), we can indeed turn the logic around to make the conjecture: Given fixed volume $V$ and fixed internal energy $U$ as constraints, massless thermal radiation (if no black hole forms) has the largest entropy amongst all possible kinds of matter. In this way, we find that a black hole is a nature-born entropy generator by way of transforming matter into thermal radiation. How cute Nature is to realise the GSL in such a delicate way! In Jacobson's approach of space-time thermodynamics \ci{jac95}, the status of the first law of thermodynamics is {\it lowered} to a more fundamental one than that of Einstein field equation. We wonder if this can also be done to the GSL?   

Obviously, it is of vital importance that Hawking radiation is massless thermal radiation. From Page's estimation \ci{page76a}, this is indeed the case for large mass black holes for which thermodynamics can be ensured making sense.

Above approach has as much {\it thermo-flavour} as we can offer. Though we left one conjecture to be verified, this can be done case by case. It is then delightful to see that Bekenstein's insight about the S\&G of thermodynamic laws could also be understood from a thermodynamic point of view.

{\it Acknowledgement} The author would like to thank an anonymous referee for helpful comments. 

%\pagebreak
%\baselineskip=2.5ex
%
%bibliography
%
%
\bn{thebibliography}{99}
\bibitem{cal85}H.B. Callen, {\it Thermodynamics and an Introduction to Thermostatistics}, 2nd edition (John Wiley \& Sons, New York, 1985)

\bibitem{hua87}K. Huang, {\it Statistical Mechanics}, 2nd edition (John Wiley \& Sons, New York, 1987)

\bibitem{haw73}S.W. Hawking and G.F.R. Ellis, {\it The Large Scale Structure of Space-time} (Cambridge University Press, Cambridge, 1973)

\bibitem{wald84}R.M. Wald, {\it General Relativity} (The University of Chicago Press, Chicago, 1984)

\bibitem{jac95}T. Jacobson, Phys. Rev. Lett. {\bf 75} (1995) 1260 (gr-qc/9504004)
%thermodynamics of space-time: The Einstein equation of state 

\bibitem{bek73b}J.D. Bekenstein, Phys. Rev. D{\bf 7} (1973) 2333
%black holes and entropy

\bibitem{haw75}S.W. Hawking, Commun. Math. Phys. {\bf 43} (1975) 199
% Particle creation by black holes

\bibitem{bar73}
J.M. Bardeen, B. Carter, and S.W. Hawking, Commun. Math. Phys. {\bf 31} (1973) 161
%{The four laws of black hole mechanics}

\bibitem{bek74a}J.D. Bekenstein, Phys. Rev. D{\bf 9} (1974) 3292
%{Generalised second law of thermodynamics in black hole physics}

\bibitem{unr82}W.G. Unruh and R.M. Wald, Phys. Rev. D{\bf 25} (1982) 942
%Acceleration radiation and the generalised second law of %thermodynamics

\bibitem{zur82}W.H. Zurek, Phys. Rev. Lett. {\bf 49} (1982) 1683
%Entropy evaporated by a black hole

\bibitem{sch85}B.W. Schumacher, Phys. Rev. Lett. {\bf 54} (1985) 2643
%Is Bekenstein's conjecture true for charged black holes

\bibitem{zur85}W.H. Zurek and K.S. Thorne, Phys. Rev. Lett. {\bf 54} (1985) 2171
%Statisticla mechanics origin of the entropy of a rotating, %charged black hole

\bibitem{fro93}V.P. Frolov and D.N. Page, Phys. Rev. Lett. {\bf 71} (1993) 3902 (gr-qc/9302017)
%proof of the generalised second law for quasistationary %semiclassical black %hole GR-QC

\bibitem{zas96}O.B. Zaslavskii, Class. Quant. Grav. {\bf 13} (1996) L7
%Generalized second law and the Bekenstein,s bound in Gedankenexperiments with %black holes

\bibitem{bek94}J.D. Bekenstein, Phys. Rev. D{\bf 49} (1994) 1912 (gr-qc/9307035)
%Entropy bounds and black hole remnants

\bibitem{sung97a}S.-T. Sung, {\it A Quantum Material Model of Static Schwarzschild Black Holes} (gr-qc/9703039)

\bibitem{haw76a}S.W. Hawking, Phys. Rev. D{\bf 13} (1976) 191
% Black hole and thermodynamics  

\bibitem{cho93}M.W. Choptuik, Phys. Rev. Lett. {\bf 70} (1993) 9
%Universality and scaling in gravitational collapse of a massless scalar field

\bibitem{abr93}A.M. Abrahams and C.R. Evans, Phys. Rev. Lett. {\bf 70} (1993) 2980
%Critical behavior and scaling in vacuum axisymmetric gravitational collapse

\bibitem{eva94}C.R. Evans and J.S. Coleman, Phys. Rev. Lett. {\bf 72} (1994) 1782
%Critical behavior and scaling in vacuum axisymmetric gravitational collapse

\bibitem{tol34}R.C. Tolman, {\it Relativity, Thermodynamics, and Cosmology} (Oxford University Press, London, 1934)

\bibitem{gib78a}
G.W. Gibbons and M.J. Perry, Proc. Roy. Soc. Lond. A{\bf 358} (1978) 467
% Black holes and thermal Green's function

\bibitem{page76a}D.N. Page, Phys. Rev. D{\bf 13} (1976) 198
% Particle emission rates from a black hole: massless particles from an %uncharged, nonrotating hole

\end{thebibliography}
\end{document}